\documentclass{article}
\usepackage[utf8]{inputenc}
\usepackage{textgreek}
\DeclareUnicodeCharacter{2212}{\ensuremath{-}}

\usepackage[english]{babel}
\usepackage{comment}

\usepackage[a4paper,top=2cm,bottom=2cm,left=3cm,right=3cm,marginparwidth=1.75cm]{geometry}

\usepackage{amsmath}
\usepackage{graphicx}
\usepackage[colorlinks=true, allcolors=blue]{hyperref}
\usepackage{subfigure}
\usepackage{float}
\usepackage{cite}
\usepackage{adjustbox} 
\usepackage{graphicx} 
\usepackage{geometry} 
\usepackage{caption} 
\usepackage{multicol} 
\usepackage{chemfig}
\usepackage{array}
\usepackage{booktabs}
\title{Chemical Origin of Exciton Self-trapping in Cs$_3$Cu$_2$X$_5$ Cesium Copper Halides}
\author{Zijin Wu$^{\dag}$, Geert Brocks$^{\dag, \ddag}$, and Shuxia Tao$^{*,\dag}$}

\begin{document}
\maketitle
\begin{center}
    $\dag$: Department of Applied Physics and Science Education, Eindhoven University of Technology, 5600 MB, Eindhoven, The Netherlands\\ 

    $\ddag$: Computational Chemical Physics, Faculty of Science and Technology and MESA+ Institute for Nanotechnology, University of Twente, 7500 AE, Enschede, The Netherlands\\

    \vspace{0.5 cm} 
    *E-mail: s.x.tao@tue.nl
\end{center}

\begin{abstract}
Copper halides Cs$_3$Cu$_2$X$_5$ (X = Cl, Br, I) are promising materials for optoelectronic applications due to their high photoluminescence efficiency, stability, and large Stokes shifts. In this work, we uncover the chemical bonding origin of the Stokes shift in these materials using density functional theory calculations. Upon excitation, one [Cu$_2$X$_5$]$^{3-}$ anion undergoes sizeable local distortions, driven by Cu–X and Cu–Cu  bond formation. These structural changes coincide with the formation of a self-trapped exciton, where particularly the hole is strongly localized on one anion. Analysis of the electronic structure and bonding reveals reduced antibonding interactions and enhanced bonding character in the excited state, stabilizing the distorted geometry. Our results establish a direct link between orbital-specific hole localization and bond formation. It provides a fundamental understanding of the excitation mechanism in Cs$_3$Cu$_2$X$_5$ and offers design principles to tune optical properties in 0D copper halides.
\end{abstract}

Cesium copper halides with the general formula Cs$_3$Cu$_2$X$_5$ (X = I, Br, Cl) are promising, nontoxic, materials for optoelectronic applications. Single crystals of Cs$_3$Cu$_2$I$_5$ exhibit photoluminescence quantum yields (PLQY) exceeding 90\% at room temperature \cite{jun}. By substituting iodide by bromide or chloride, the emitted light can be tuned from blue to green \cite{Y.Li}. The compounds demonstrate considerable stability, retaining their luminescence for extended periods in air and under operation \cite{Z.Luo}. Furthermore, even after degradation, photoluminescence can be restored by re-annealing the material \cite{H.Ding}. This reversible luminescent behavior makes Cs$_3$Cu$_2$X$_5$ materials particularly attractive for anti-counterfeiting applications\cite{f.gao}. 

The most distinctive optical property of Cs$_3$Cu$_2$X$_5$ compounds is the wavelength shift from the photoluminescence excitation (PLE) peak to the photoluminescence (PL) emission peak, known as the Stokes shift, which is extraordinarily large. Depending on the halide composition, the Stokes shift in these materials ranges from 150 to 250 nm \cite{Z.Luo}, resulting in a near-zero spectral overlap between PLE and PL peaks. The lack of absorption in the optical region where emission takes place, makes these materials of interest for application in high-resolution, fast-response, and ultra-sensitive radiation detectors \cite{Lian}. Tuning the halide composition enables to adjust the Stokes shift, which, together with the high PLQY, further enhances the potential of Cs$_3$Cu$_2$X$_5$ materials \cite{B.Yang}.

The large Stokes shift found in these materials is attributed to self-trapped excitons (STEs) \cite{jun_t}.Copper halide-based compounds frequently display STEs. For instance, STE electronic configurations and associated geometry distortions have also been characterized in 1D cesium copper halides CsCu$_2$X$_3$ \cite{du_m}. The Cs$_3$Cu$_2$X$_5$ crystal structure comprises discrete [Cu$_2$X$_5$]$^{3-}$ anions, separated by Cs$^+$ cations. In the classification of metal halides this is dubbed a zero-dimensional (0D) material \cite{Arslanova2024}. Lian \emph{et al.} proposed that the lowest energy electronic excitation is localized on one [Cu$_2$X$_5$]$^{3-}$ ion, which causes a large geometry distortion of that ion \cite{Lian}. The Stokes shift corresponds to the energy difference between optical absorption from the ground-state equilibrium geometry and photoluminescence emission from the relaxed excited-state geometry \cite{Lian}.  

Whereas the presence of substantial geometry distortions upon electronic excitations in copper halides is well argued, the chemical driving force behind a particular distortion is often elusive. In particular, in 0D copper halides excitation drives a [Cu$_2$X$_5$]$^{3-}$ ion toward a more compact and more symmetric geometry \cite{Lian}, which could be interpreted as a sign of increased chemical bonding. This is somewhat peculiar, as excitation is typically associated with bond breaking and a decrease in chemical bonding. In the present paper we explore the chemical origin of charge and exciton localization in Cs$_3$Cu$_2$X$_5$ compounds, using density functional theory (DFT) calculations. 

We argue that localization of a hole on a single [Cu$_2$X$_5$]$^{3-}$ ion is the driving force. This process takes an electron away from an antibonding orbital, which increases bonding and pushes the geometry toward a more compact and symmetric configuration. The charge localization and concomitant geometry transformation are identified from hole-only DFT+U calculations. Building upon that, very similar patterns are found in $\Delta$SCF calculations on singlet and triplet excitations, confirming that hole localization is the essential driving force for increased chemical bonding in the excited state, promoting a compact structure. The substantial geometry change induced by the change in chemical bonding is the origin of the large Stokes shifts observed in these materials.

The chemistry of these copper halides is consistent with what is commonly found in copper d$^{10}$ compounds, where there is a subtle balance between bonding and anti-bonding states involving the Cu d-orbitals \cite{mehrotra,MerzIC1988}. This balance may be easily disrupted by excitations, suggesting that such compounds are ideally suited for promoting large Stokes shifts, and make them suitable for PL applications.

    \begin{figure}[htb]
        \centering
        \includegraphics[width=1.0\textwidth]{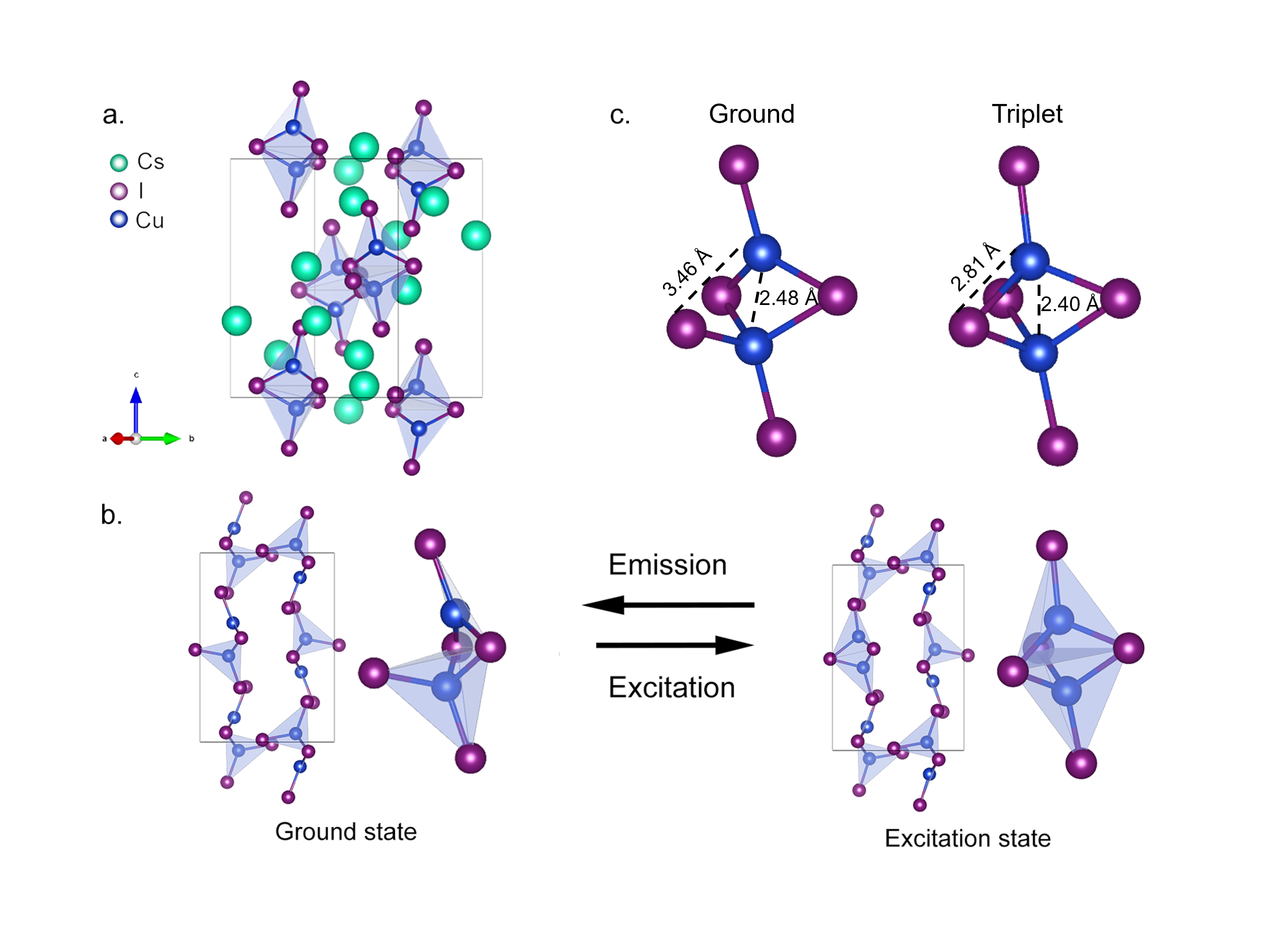} 
        \caption{\textbf{a.} Cs$_3$Cu$_2$I$_5$ ground state structure. \textbf{b.} Proposed geometry change of [Cu\textsubscript{2}I\textsubscript{5}]\textsuperscript{3-} ion upon excitation. The Cs ions are hidden to highlight the structural change. \textbf{c.} Most prominent changes in bond lengths in [Cu\textsubscript{2}I\textsubscript{5}]\textsuperscript{3-} upon excitation.} 
        \label{geo} 
    \end{figure}

\emph{Structures:} In their ground state, Cs$_3$Cu$_2$X$_5$ (X = I, Br, Cl) compounds adopt an orthorhombic crystal structure with space group \textit{Pnma} \cite{z.jiang}. We start from experimental structures \cite{z.jiang}, which are subsequently optimized in DFT calculations. Technical details of these calculations can be found in the Methods section; the optimized structures are given in the Supporting Information (SI), Figs. S2, S3. Within the unit cell, Cu and X ions form four [Cu$_2$X$_5$]$^{3-}$ anions, while Cs$^+$ cations primarily act as spacers, separating the Cu$_2$X$_5$ units, as illustrated in Fig.\ref{geo}(a). 

The structure of a single [Cu$_2$X$_5$]$^{3-}$ anion is asymmetric. It consists of a triangular CuX$_3$ structure and a tetrahedral CuX$_4$ structure, connected through a shared edge, see Fig.~\ref{geo}(b). This structure may seem somewhat unusual, but it is consistent with the structural patterns found in other copper(I) halides \cite{PengCCR2010}. Halide ions can function as terminal ligands, but also as bridging ligands, where with copper(I) ions they can form chain-like structures --X--Cu--X--Cu-- \cite{A902290B}. A chain can close upon itself, and the smallest unit formed this way is a rhombus, \[\chemfig{X*4(-Cu-X-Cu-)}\] which is a building block that is very often found in mixed organic copper halides \cite{PengCCR2010, A902290B, Blake1998}. Linking such rhombi to larger structures via the Cu sites is a possibility \cite{A902290B, Blake1998}, but attaching one or two terminal halides to the Cu sites and isolate a single rhombus is also common \cite{CaradocDalton2002}. 

Terminating one Cu site by a single atom, and the other by two, as in, \[\chemfig{X*4(-Cu(<|[6]X)(>|[8]X)-X-Cu(-[3]X)-)}\] is basically the structure we find for the [Cu$_2$X$_5$]$^{3-}$ anions. This configuration is somewhat atypical, but not unique for the current case \cite{CaradocDalton2002}. Copper(I) halide structures tend not to be extremely rigid, as Cu ions in copper(I) halides have completely filled 3d shells and their 4p states have quite a high energy, so bonding is not extremely directional. Details of the structure, bonding angles in particular, are then often determined by the packing with other ions/ligands \cite{PengCCR2010,CaradocDalton2002}, which in the present case leads to the asymmetric [Cu$_2$X$_5$]$^{3-}$ structure. 

Lian \emph{et al.} argue that the excited states of the Cs$_3$Cu$_2$X$_5$ crystal are localized on individual Cu$_2$X$_5$ units, and that, upon excitation a Cu$_2$X$_5$ unit attains a more symmetric structure \cite{LianCM2020}. That structure is closer to a confacial bitetrahedral complex, where each of the two Cu ions is tetrahedrally surrounded by halides, and the two tetrahedrons share a face, see Fig.~\ref{geo}(b). The structural change of the Cu$_2$X$_5$ unit is thought to be at the basis of the large Stokes shift, and hence central to the functionality of the Cs$_3$Cu$_2$X$_5$ compounds as luminescent materials \cite{LianCM2020}. 

Confacial bitetrahedral structures are found in particular mixed organic inorganic compounds that involves Cu$_2$Br$_5$ clusters, where it is speculated that these structures are stabilized by a Cu-Cu bond \cite{HornJACS1998}. For the present compounds that would be extraordinary, as such a bond formation would be achieved upon excitation of the Cu$_2$X$_5$ unit. One important feature of the [Cu$_2$Br$_5$] anion in the compound or Ref. \cite{HornJACS1998} is its charge state of $2-$, whereas in Cs$_3$Cu$_2$Br$_5$ it is $3-$. This suggests that excitation might effectively change the anion's charge state, which then might drive a geometry change.

A scheme of the excitation–emission cycle is provided in Fig.~\ref{geo}(b). In the excited state, the most significant structural change occurs within one of the [Cu$_2$X$_5$]$^{3-}$ ions as a transformation from the edge-sharing tetrahedron–triangle unit to a bitetrahedral structure. From the structural change one can deduce that an additional Cu-X bond is formed, and the Cu-Cu interaction is strengthened. This is depicted in Fig.~\ref{geo}(c) for Cs$_3$Cu$_2$I$_5$, where upon excitation one Cu-I distance is shortened by 0.7 \AA\ to a value that is common for Cu-I bonds. At the same time, the Cu-Cu distance is shortened by 0.1 \AA. The same patterns are found in the other Cs$_3$Cu$_2$X$_5$ compounds, see the SI, Figs. S2-S5. After emission the bitetrahedral [Cu$_2$X$_5$]$^{3-}$ cluster relaxes back to its original asymmetric edge-sharing tetrahedron–triangle shape, making the process reversible. 

\begin{figure}[htb]
    \centering
    \includegraphics[width=1\linewidth]{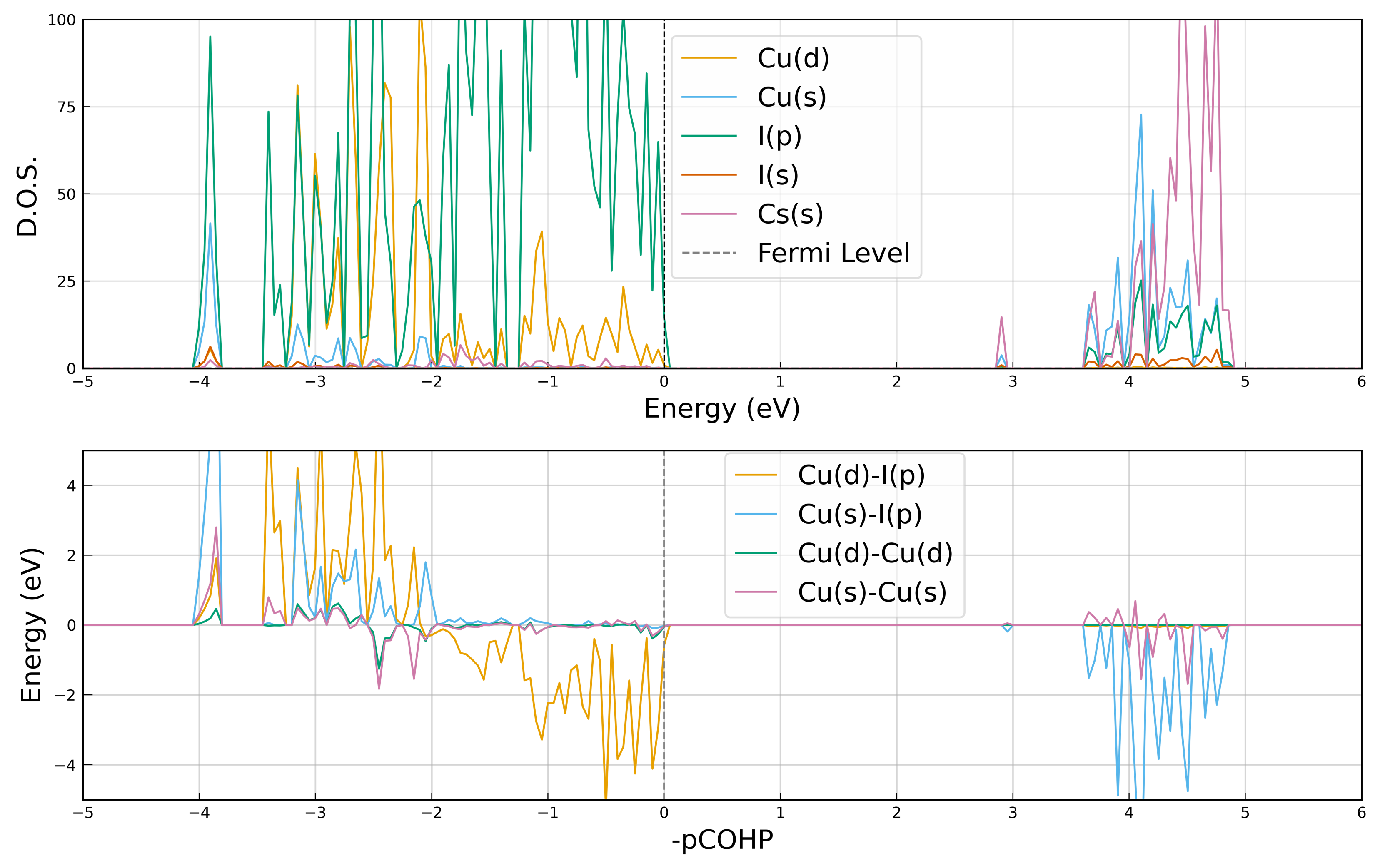} 
    \caption{Top: density of states (DOS) and bottom: crystal orbital Hamilton population (COHP) of Cs$_3$Cu$_2$I$_5$ in the ground state. The zero level is put at the top of the valence band and is marked by a  dashed line.}
    \label{dos} 
\end{figure}

\emph{Bonding:} The bonding in Cs$_3$Cu$_2$X$_5$ compounds can be analyzed by examining the density of states (DOS) and the crystal orbital Hamilton population (COHP). The results for the ground state of Cs$_3$Cu$_2$I$_5$ are shown in Fig.~\ref{dos}, and the results for Cs$_3$Cu$_2$Br$_5$ and Cs$_3$Cu$_2$Cl$_5$ are shown in the SI, Fig.~\ref{si_dos}. The orbital resolved DOS clearly shows that the electronic structure of Cs$_3$Cu$_2$I$_5$ at the top of the valence band, between $-3.5$ and $0$ eV, is dominated by halide and copper contributions, specifically by Cu($d$) and I($p$) orbitals. The COHP analysis of states in this energy region reveals that the lower half ($-3.5$ to $-2$ eV) is formed by bonding bonding Cu($d$)--I($p$) states, whereas the upper half ($-2$ to $0$ eV) comprises antibonding Cu($d$)--I($p$) states. 

The DOS shows the bonding states to have larger Cu($d$) contributions, and the anti-bonding ones larger I($p$) contributions. Interpreting this in terms of hybridization between I($p$) and Cu($d$) atomic states, it means that the former have a higher energy than the latter. The bottom of the conduction band ($\sim 4$ eV) has large Cu($s$) contributions, and the states have antibonding Cu($s$)--I($p$) character, with a small Cu($s$)--Cu($s$) contribution. Overall, the electronic structure is consistent with atomic valencies/configurations Cu$^+3d^{10}4s^0$ and Cl$^-3s^23p^6$ before hybridization. 

The bonding analysis of Cs$_3$Cu$_2$X$_5$ gives very similar results for X$=$Cl, Br, I. The influence of halide substitution on the electronic structure is illustrated in Figs.~\ref{si_dos},\ref{si_br_cohp} and \ref{si_cl_cohp}. The upper valence bands all indicate a bonding/anti-bonding interaction between Cu($d$) and X($p$) states, and the lower conduction states have antibonding Cu($s$)--X($p$) character. For X$=$Cl, the bonding states have larger Cl($p$) contributions, and the antibonding ones have more Cu($d$) character, implying that the Cl($p$) atomic states have a lower energy than the Cu($d$) atomic states. For X$=$Br, the Cu($d$)/Br($p$) character seems to be evenly distributed over bonding and anti-bonding states, indicating that the Cu($d$) and Br($p$) atomic states have approximately the same energy. This behavior is consistent with the fact that the energy of the highest occupied X($p$) atomic shell decreases in the order I $\rightarrow$ Br $\rightarrow$ Cl. As a further consequence of that, one expects in that order the top of the valence band to decrease in energy and the band gap to increase. Fig.~\ref{si_dos} shows that that is indeed the case.

Fig.~\ref{si_ground_geo} lists the nearest-neighbor distances of atom pairs in the Cu$_2$I$_5$ clusters in their ground state. Remarkable is the rather short Cu-Cu distance of 2.5 \AA, which may be interpreted as a sign of a weak Cu-Cu bond, as proposed by Hoffmann and coworkers \cite{mehrotra,MerzIC1988}. The closed shell 3d$^{10}$ configuration of Cu$^+$ gives no motive for such bond formation, but weak Cu-Cu bonding may be driven by hybridization of 3d with 4s/4p orbitals.\cite{mehrotra,MerzIC1988}. Terminal Cu-I distances are around 2.5 \AA, which is typical for Cu-I bonds. The two Cu-I distances in the central rhombus are 2.7 \AA, consistent with the notion that bridging halogens give slightly longer bonds than terminal ones. The Cu-I distance indicated in Fig.~\ref{geo}(c) is 3.5 \AA, which is considerably larger than the typical Cu-I bonding distance, and thus signifies the absence of a bond between that Cu and I. 

Comparing Cs$_3$Cu$_2$X$_5$ compounds in the Cl~$\rightarrow$~Br~$\rightarrow$~I series one observes a systematic lattice expansion that can be attributed to the increasing size of the halide atom, which results in a progressive lengthening of the Cu--X bonds. Relative to Cu--Cl, the Cu--Br bond is longer by \mbox{$\approx 0.14$~\AA}, and the Cu--I bond by \mbox{$\approx 0.30$~\AA} (Table.~\ref{ground_bond}).  Accordingly, the Br- and I-containing compounds display larger lattice parameters; for example, the $a$-axis length of Cs$_3$Cu$_2$I$_5$ (10.08~\AA) is larger than that of Cs$_3$Cu$_2$Cl$_5$ by nearly 1~\AA. In contrast, the Cu--Cu bond length remains essentially unchanged across the series, increasing by at most $\sim$0.03~\AA.  This negligible change indicates that halide substitution has little influence on the Cu--Cu interaction. In all cases the identified Cu-X distance is $\sim 1$ \AA\ larger than a typical Cu-X bonding distance, indicating the absence of a bond.

\begin{figure}[H]
    \centering
    \includegraphics[width=0.6\linewidth]{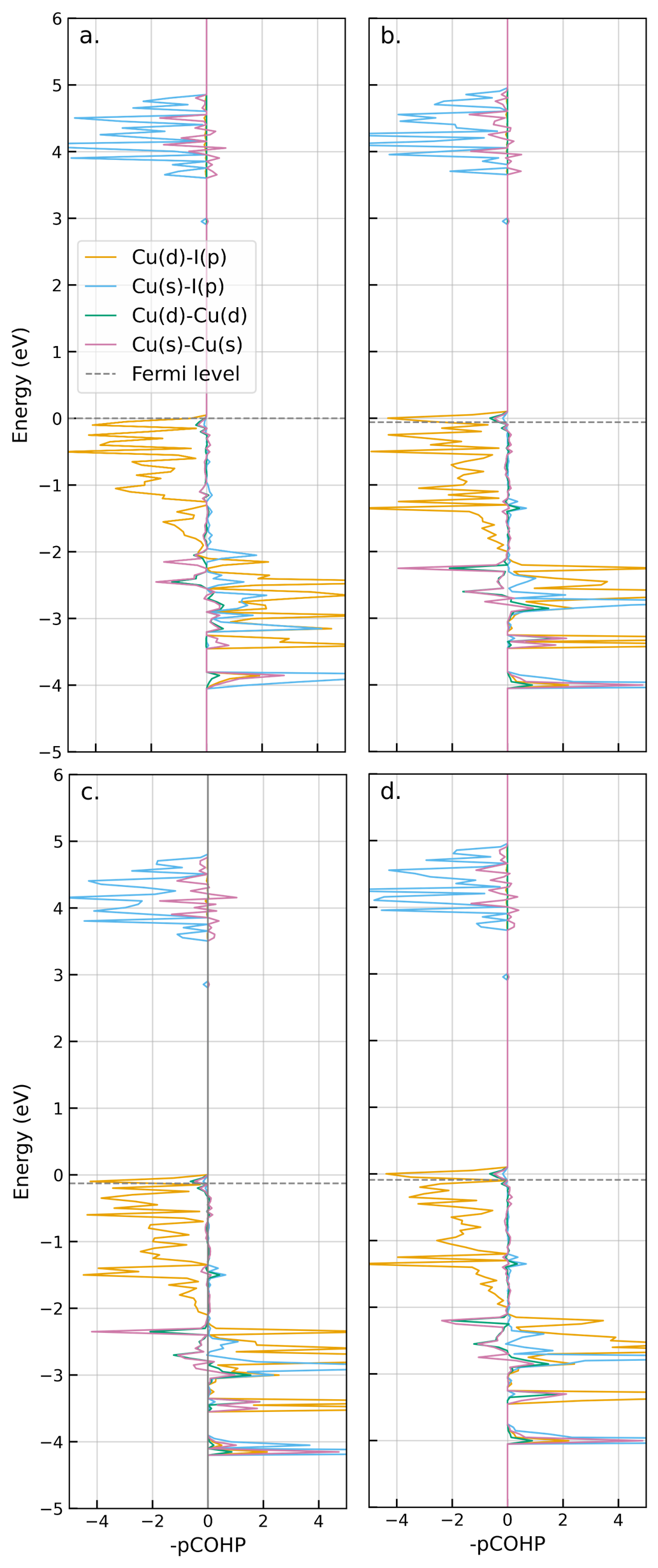} 
    \caption{COHP of the affected [Cu\textsubscript{2}I\textsubscript{5}]\textsuperscript{3-} unit in Cs$_3$Cu$_2$I$_5$ in (a) the ground state, (b) one hole added, and the hole channels of the (c) singlet and (d)triplet states. The Fermi level is marked by the horizonal dash line.}
    \label{cohp} 
\end{figure}

\emph{Adding holes:} As discussed above, [Cu$_2$Br$_5$]$^{2-}$ ions can be found in a confacial bitetrahedral structure \cite{HornJACS1998}. For [Cu$_2$X$_5$]$^{3-}$ ions the more frequently occurring structure is the more open one, based upon a [CuX]$_2$ rhombus \cite{PengCCR2010,CaradocDalton2002}, Fig.~\ref{geo}(c). From this observation one may speculate that the geometry of a Cu$_2$X$_5$ ion is linked to its charge. The individual bands at the top of the valence band of Cs$_3$Cu$_2$X$_5$ have a very small dispersion ($\sim 0.1$ eV), indicating that states are localized on the Cu$_2$X$_5$ ions, see Fig.~\ref{i_band}. Adding a hole to the valence band, that hole may possibly localize on one Cu$_2$X$_5$ ion, changing its charge from $3-$ to $2-$.  

In contrast, the dispersion of bands at the bottom of the conduction band is at least an order of magnitude larger, signaling that states are more delocalized. An electron added to the conduction band is less likely to localize, suggesting that exciton self-trapping is mainly driven by hole localization. That suggestion would be in line with scenarios proposed for exciton self-trapping in alkali and alkaline earth metal halides, which also rely upon hole localization \cite{WilliamsJPCS1990}. It therefore makes sense to start the analysis of exciton self-trapping by studying what happens if holes are added to the system.    

Optimizing the ground state structure of a $2\times 2\times 1$ supercell of Cs$_3$Cu$_2$X$_5$ with one hole added (one electron removed), we find that indeed the hole localizes on one Cu$_2$X$_5$ unit, which then changes its geometry from open rhombus type to bitetrahedral, with a relaxation energy of 0.2 eV. The COHP of the Cu$_2$I$_5$ unit with one hole added is shown in Fig. \ref{cohp}(b). The topmost state of the valence band, which is now occupied by the hole, has a dominant Cu($d$)--I($p$) antibonding character, with a small Cu($s,d$)-Cu($s,d$) antibonding contribution. In addition, we find that the overall width of the Cu($d$)--I($p$) bonding/antibonding band has increased going from the pristine [Cu$_2$X$_5$]$^{3-}$ to the geometry optimized [Cu$_2$X$_5$]$^{2-}$ ion, Figs. \ref{cohp}(a)(b), indicating a stronger hybridization in the latter.   

\begin{figure}[H]
    \centering
    \includegraphics[width=0.5\linewidth]{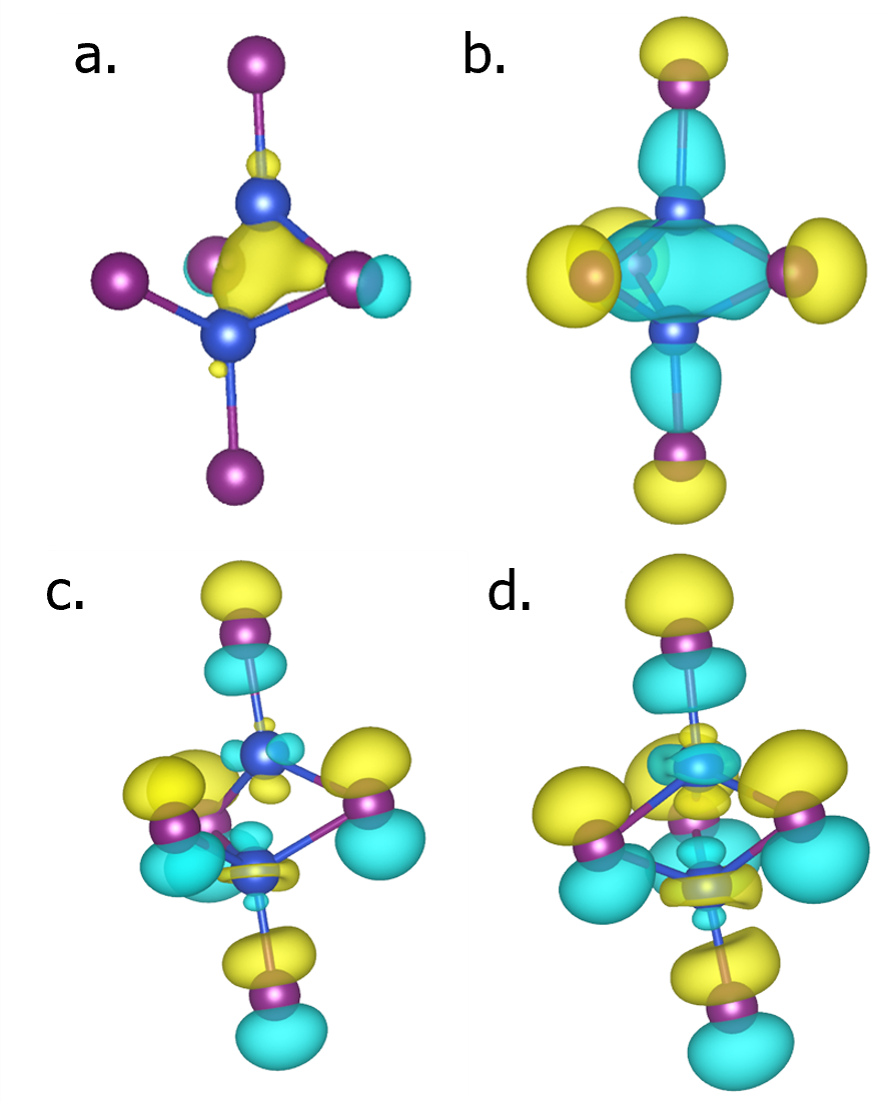} 
    \caption{Wave functions of the lowest Cu($d$)--I($p$) bonding state (a,b) and the highest antibonding state (c,d) of the [Cu$_2$I$_5$]$^{2-}$ ion. Yellow and blue colors denote regions of opposite phase; (a,c) undistorted [Cu$_2$I$_5$]$^{3-}$ rhombus structure; (b,d) optimized [Cu$_2$I$_5$]$^{2-}$ bitetrahedral structure.}
    \label{wavefunc} 
\end{figure}

In the pristine state, Fig. \ref{cohp}(a), all Cu($d$)--I($p$) bonding and antibonding states are occupied, which implies that Cu($d$)--I($p$) hybridization does not contribute to any net bonding. This would make sense, as in a [Cu$_2$Cl$_5$]$^{3-}$ cluster both Cu ions have completely filled $d^{10}$ shells. Emptying the topmost antibonding level in one spin channel, Fig. \ref{cohp}(b), tops the balance in favor of the Cu($d$)--I($p$) bonding states, and leads to strengthening the bonding. We get an idea of the bonding by examining the lowest Cu($d$)--I($p$) bonding state (at $\sim -3.5$ eV in Fig. \ref{cohp}(a,b)) and the highest antibonding state (at $\sim 0$ eV in Fig. \ref{cohp}(a,b)) in the ground state structure and in the bitetrahedral structure. The corresponding wave functions are shown in Fig. \ref{wavefunc}. 

The lowest bonding state involves $\sigma$-bonding between Cu($d_{z^2}, d_{x^2-y^2}, d_{xy}$) hybrids and the I($p$) orbitals, with some contribution from a direct $\sigma$-interaction between $d$-orbitals on the two Cu atoms. In the ground state structure the bonding state involves the two iodines in the [CuI]$_2$ rhombus. In the bitetrahedral structure all three iodine atoms in the $xy$-plane are involved in the bonding state, creating a larger bonding overlap. Indeed, in the bitetrahedral structure this bonding state is then found at a lower energy.

The highest anti-bonding state can be characterized as $\sigma^*$ between the $d$-orbitals on the two Cu atoms, as $\sigma^*$ between the Cu($d$) and I($p$) of the axial I atoms, and as $\pi^*$ between the Cu($d$) and I($p$) of the I atoms in the $xy$ plane. Again, in the ground state structure only two I atoms are involved, whereas in the bitetrahedral structure the three I atoms in the $xy$-plane participate. This makes the latter state more anti-bonding and thus found at a higher energy. If this state is depopulated by introducing a hole, the bitetrahedral structure is stabilized by allowing it to benefit from the larger bonding overlap.  

This reasoning also agrees with very general chemical considerations. If we formally assign the hole to one Cu atom, it would locally create a $3d^9$ configuration with oxidation state 2+. As Cu$^{2+}$ is a stronger Lewis acid than Cu$^+$, it prefers to have a larger coordination number \cite{PavelkaCP2005}, which leads to a more compact Cu$_2$I$_5$ cluster. 

\emph{Excitations:} The lowest triplet excitation is accessible by Kohn-Sham DFT calculations, as it is the ground state of the triplet manifold \cite{JonesRMP1989}. To model the triplet, we perform an unrestricted self-consistent field (SCF) calculation while populating the spin-up states by two more electrons than the spin-down states. The singlet exciton can formally not be captured by ground state DFT, as it represents an excited state \cite{JonesRMP1989}. Using a pragmatic approach, one can manually transfer an electron from an occupied to an unoccupied orbital, followed by self-consistent field (SCF) optimization while freezing these non-Aufbau occupations of the orbitals \cite{HellmannJCP2004,ChengJCP2008,KowalczykJCP2011,HaitJPCL2021}. This state presents a stationary point to which SCF optimization can converge, provided the characteristics of populated and depopulated orbitals are sufficiently different. 

Expressing the excitation energy as the difference between the total energy of the (singlet or triplet) excited state and that of the ground state, this method is coined $\Delta$SCF. Although spin contamination can hamper the accuracy of unrestricted $\Delta$SCF, calling for a spin purification procedure \cite{KowalczykJCP2011}, in the current cases the singlet-triplet splitting energy turns out to be small ($<0.1$ eV), which makes such a procedure unnecessary.  

$\Delta$SCF calculations can be combined with geometry optimizations. We explored ground state and bitetrahedral geometries for the [Cu$_2$X$_5$]$^{3-}$ ions. Both in triplet and singlet excited states, the system favors a structure characterized by the transformation of one of the [Cu$_2$X$_5$]$^{3-}$ ions in the supercell into a bitetrahedral structure, while the other ions preserve their original structure. The calculated relaxation energies are $\sim 0.2$ eV, with negligible differences between singlet and triplet states. The differences between the ground state and bitetrahedral structures are also extremely similar to the hole-only case discussed above.

Comparing the optimized geometries of Cs$_3$Cu$_2$X$_5$, X = Cl, Br, I, compounds we observe that in the optimized singlet and triplet geometries the Cu--Cu bond length is about 0.1 \AA\ shorter than in the ground state geometry for all three compounds, see SI Table.\ref{ground_bond} and Table.\ref{triplet_bond}. The most prominent change going from the ground state to the excited state geometry is the formation of a new Cu-X bond within one Cu$_2$X$_5$ ion.
The change in the corresponding Cu--X distance is similar for all three compounds, i.e., 0.6--0.7 \AA, whereas the other Cu--X bonds did not exhibit any substantial changes. These geometry changes are very similar to what has been found for the added-hole systems, which is a clear indication that the localization of the hole is the driving force for the geometry change. 

\begin{figure}[htb]
    \centering
    \includegraphics[width=1\linewidth]{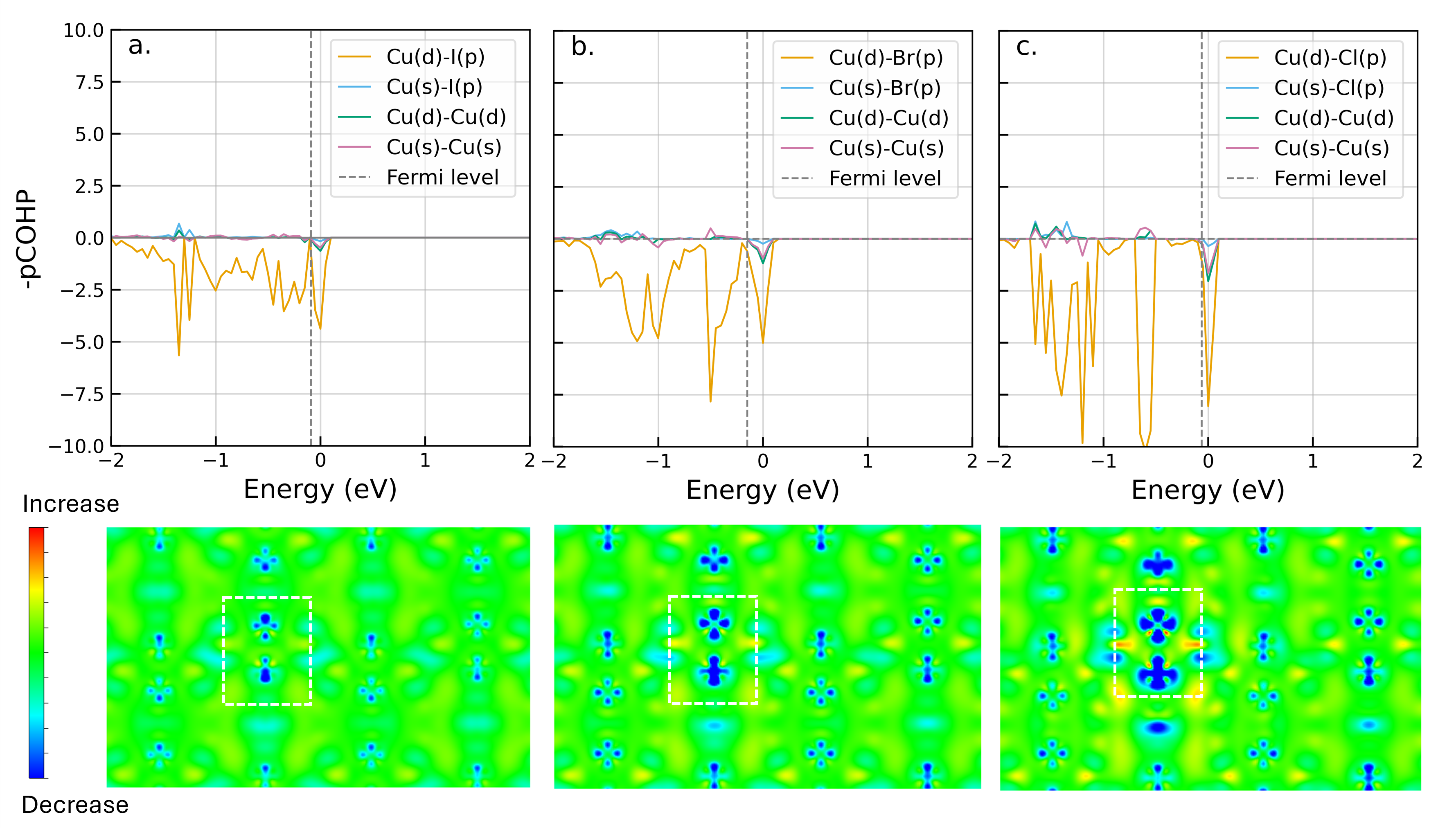}
    \caption{COHP of hole spin channel of the triplet (top) and $\Delta n(\mathbf r) = n_\mathrm{triplet}(\mathbf r) - n_0(\mathbf r)$ (bottom) of Cs$_3$Cu$_2$X$_5$, X = (a) I, (b) Br, (c) Cl. The Cu$_2$X$_5$ ion with bitetrahedral geometry is marked by the white frame.}
    \label{cdd}
\end{figure}

The electronic structures of triplet and singlet also show a high similarity to the hole-only case. As an example, Fig. \ref{cohp} compares COHPs  of the hole channels of triplet, singlet and hole-only Cs$_3$Cu$_2$I$_5$. Their similarity is a clear indication of the localization of the hole on one Cu$_2$I$_5$ ion in all cases. All halide variants of Cs$_3$Cu$_2$X$_5$ exhibit similar features, including hole localization and a corresponding emptying of the anti-bonding state near the fermi level, as illustrated in Fig.~\ref{cdd}. It stands to reason that the proposed stabilization mechanism for the bitetrahedral structure of the excited states is also the same as for the hole-only case. Hybridization between Cu($d$) and X($p$) states in the Cu$_2$X$_5$ core is larger in the bitetrahedral geometry, Fig. \ref{wavefunc}, resulting in a larger bonding/antibonding splitting. Localization of the hole empties the topmost antibonding state, thereby strengthening the bonding.

To visualize the spatial distribution of the exciton, we analyze the electron density difference relative to the ground state, $\Delta n(\mathbf r) = n_\mathrm{exc}(\mathbf r) - n_0(\mathbf r)$. In this calculation we fix the structure of one of the Cu$_2$X$_5$ ions at the bitetrahedral geometry of the excited state. Fig.~\ref{cdd} shows 
two-dimensional cuts through $\Delta n(\mathbf r)$ for the triplet in Cs$_3$Cu$_2$X$_5$, X = I, Br, Cl. Identifying negative (positive) $\Delta n(\mathbf r)$ values with the hole (electron) of the excitation reveals that the hole largely localizes on the bitetrahedral cluster, whereas the electron is more delocalized. 

The shape and orientation of the hole localization region suggest that important contributions come from Cu($d_{z^2},d_{x^2-y^2}$) orbitals and X($p_z$) orbitals of the halogens in the equatorial plane, consistent with our analysis of the antibonding state, Fig. \ref{wavefunc}(d). Additionally, an increase of electron density is observed in the core region of the bitetrahedral cluster, which can be interpreted as an increase of the Cu-X bonding in that region.  

Comparing $\Delta n(\mathbf r)$ for the different compounds Cs$_3$Cu$_2$X$_5$, X = Cl, Br, I, one observes that the Cl-based copper halide displays the strongest hole localization, followed by the Br-based compound, while the I-based Cs$_3$Cu$_2$I$_5$ shows the weakest localization. This trend is in agreement with the variations observed in the COHP, where the peak corresponding to the antibonding state that is emptied upon excitation, is sharpest and most isolated in Cs$_3$Cu$_2$Cl$_5$, which is indicative of a very localized state.

We attribute this to the fact that the X$(p)$ states decrease in energy in the order I $\rightarrow$ Br $\rightarrow$ Cl, which implies that the antibonding states at the top of the valence band have increasing Cu$(d)$ character, and are thus more localized. In addition, as Cu-X distances decrease in the same order, that deepens the potential well by enhancing the Coulombic interaction, thereby promoting the stronger hole localization in the Cl-based system.

The $\Delta$SCF excitation energies without relaxation ($\Delta$SCF$_0$) are 2.8 eV, 3.0 eV, and 3.4 eV for Cs$_3$Cu$_2$X$_5$, X = I, Br, Cl, respectively. The energy difference between singlet and triplet excitations is remarkably small ($\sim 0.1$ eV). This may be explained by the fact that the only difference between the two (in $\Delta$SCF) is the spin state of the electron and as the electron is only weakly localized, the electron-hole exchange interaction is not expected to be very strong. The calculated relaxation energy is 0.2 eV with only small differences between the compounds. Approximating the Stokes shift by twice the relaxation energy gives it a value of $\sim 0.4$ eV.

\emph{Discussion:} The photoluminescence (PL) of Cs$_3$Cu$_2$X$_5$ compounds is fairly robust under variation of synthesis conditions and materials state (poly/nanocrystals, powders, thin films, or single crystals), with PL maxima found in relatively narrow energy windows, 2.72-2.84 eV for X = I, 2.67-2.72 eV for X = Br, and 2.35-2.43 eV for X = Cl \cite{LiMCF2021,ZhangFOE2021,QuJMCC2023}. In contrast, there is a much larger variation in photoluminescence excitation (PLE) spectra, with PLE maxima found in energy windows ranging from 3.58-4.35 eV for X = I, 3.80-4.61 eV for X = Br, to 3.58-4.78 eV for X = Cl \cite{ZhangFOE2021,QuJMCC2023}.  Correspondingly, the reported Stokes shifts span quite a large energy range, from 0.76-1.56 eV for X = I, 1.14-1.92 eV for X = Br, to 1.15-2.43 eV for X = Cl \cite{ZhangFOE2021,QuJMCC2023}.  

On the basis of the reported PLE maxima it is not so clear what the chemical trend in the Cs$_3$Cu$_2$X$_5$, X = I, Br, Cl compounds is. The general tendency seems to be that the PLE maximum for X = Br is 0.1-0.2 eV higher than that for X = I, and the PLE maximum of X = Cl is 0-0.1 eV lower \cite{Lian,LunaCM2019,RoccanovaAEM2019}. In contrast, the chemical trend in the PL maxima is quite evident, with X = Br lower that X = I by $\sim 0.1$ eV, and X = Cl lower by $\sim 0.4$ eV. 

The $\Delta$SCF$_0$ excitation energies should approximately correspond to the PLE maxima, where we find some underestimation, which is quite usual in DFT calculations. The Stokes shifts, calculated on the basis of the relaxation energies, are at least a factor of two too small, and show no clear chemical trend. Local excitations on a Cu$_2$X$_5$ ion may involve electron correlations, which affect the amount of relaxation and are missed in a DFT calculation. Nevertheless, we claim that the trend in the calculated results is sufficiently clear to reveal the chemical origin of exciton self-trapping in these compounds.

Previous DFT calculations typically predict band gaps and $\Delta$SCF$_0$ excitation energies with the chemical trend I $<$ Br $<$ Cl, which is also what we find, where the exact numerical values depend on the functional used in the calculations \cite{Lian,RoccanovaAEM2019,WangNL2020,ZhuAFM2024}. The geometry distortions upon excitation found  from $\Delta$SCF calculations also depend somewhat on the functional used, with GGA+U and range-separated hybrid functionals giving a compact bitetrahedral geometry \cite{Lian}, as in Fig. \ref{geo}, and non-range-separated hybrid functionals preferring a more open geometry \cite{RoccanovaAEM2019,WangNL2020}. We suggest that the chemical driving force for geometry change is independent of functional, i.e., localization of a hole on a Cu$_2$X$_5$ ion and concomitant emptying of an anti-bonding orbital. The compact geometry then reflects increased bonding by strengthening the Cu-X bonds, whereas the more open geometry focuses on strengthening the Cu-Cu bond.

\emph{Conclusion:} In conclusion, we have uncovered the chemical origin of the pronounced Stokes shift in 0D copper alkali halides, Cs$_3$Cu$_2$X$_5$, X = I, Br, Cl, by theoretical modeling based upon DFT calculations. Optical excitation leads to self-localization of the exciton on a [Cu$_2$X$_5$]$^{3-}$ anion, where in particular the hole is strongly localized on one anion. Analysis of the electronic structure and bonding reveals that the hole occupies an antibonding Cu($d$)-X($p$) orbital. Reduced antibonding enhances bonding, which drives a geometry distortion strengthening Cu–X and Cu–Cu bonds. The sizeable local structural reorganization, characterized by shortened Cu--Cu and Cu--X bonds, provides the key mechanism underlying the large Stokes shift. 

Our results provide an understanding of the excitation process in Cs$_3$Cu$_2$X$_5$, emphasizing the crucial role of orbital-specific hole localization in exciton self-trapping and bond formation leading to structural change. The insights obtained here offer a fundamental framework for rational design and property tuning in 0D copper halides. Although this study focuses primarily on the electronic effects in specific Cs$_3$Cu$_2$X$_5$ materials, future investigations could systematically explore the role of metal cations (by, e.g., substitutions of Cu or Cs ), as well as external perturbations such as pressure or temperature. Such studies may unlock new pathways for controlling luminescence efficiency, emission wavelength, and Stokes shifts, with potential implications for photonic and optoelectronic device applications.

\emph{Methods:} All calculations were performed within the framework of Density Functional Theory (DFT) using the Vienna Ab initio Simulation Package (VASP) \cite{vasp1}. Standard projector augmented wave (PAW) potentials were used, treating Cs $5s^25p^66s^1$, Cu $3d^{10}4^1$ and halogen $s^2p^5$ as valence electrons, and a plane wave basis with a kinetic energy cutoff of 500 eV. We used the DFT+U approach with the revised Perdew-Burke-Ernzerhof exchange-correlation functional for solids (PBEsol)\cite{PBEsol}, and the rotationally averaged Liechtenstein \emph{et al.} expression \cite{Liechtenstein} with $U-J=6$ eV on the Cu atoms.

Initial crystallographic structures of Cs$_3$Cu$_2$X$_5$ were obtained from prior literature \cite{z.jiang}. Geometry optimizations were executed with initial convergence criteria for total energy and forces of 10$^{-5}$ eV and 0.05 eV/\AA, respectively. Selected structures were further optimized using 10$^{-6}$ eV, and 0.01 eV/\AA\ as criteria. Initial structure optimizations used the unit cell containing four formula units, see SI Figs. S2-S5. Subsequently, results were confirmed in a $2\times2\times1$ supercell containing 16 formula units. The band of the occupied states show very little dispersion, see SI Figs. S6. Hence,  a $\Gamma$-point-only sampling was applied for the $2\times2\times1$ supercell.

The one-hole state was modeled by removing one electron from the neutral ground state structure, resulting in a net total charge of $+1$ per (super)cell. To facilitate the formation of the local Cu-I bond associated with the hole localization, one [Cu$_2$X$_5$]$^{3-}$ unit was initially perturbed by slightly reducing the distance between an Cu and a neighboring unbonded halogen ion. This perturbed structure was subsequently subjected to full geometry optimization. The perturbed structure has a lower energy than the unperturbed one, proving the preference for the distorted structure. 

The optimized one-hole structure was used as the starting geometry for the singlet excited state. The latter was obtained in a constrained DFT approach, i.e., by manually setting the occupation numbers, and keeping the orbital occupancies fixed during subsequent geometry optimizations. The triplet state was modeled by imposing a fixed difference between the number of spin-up and spin-down electrons, while optimizing the structure.

Electronic structures were analyzed using the (projected) density of states (P)DOS obtained from the post-processing package VASPKIT \cite{vaspkit}. Chemical bonding was characterized using the crystal orbital Hamilton population COHP analysis, as generated by LOBSTER \cite{lobster1, lobster2}. Electron density differences, wave functions and crystal structures were visualized using VESTA \cite{vesta}.

\section*{Acknowledgement}
Z.W. and S.T. acknowledge funding from Vidi (project no. VI.Vid.213.091) from the Dutch Research Council (NWO).
\bibliographystyle{unsrt}
\bibliography{sample}

\newpage
\section*{Supplementary Information}
\renewcommand{\thefigure}{S\arabic{figure}}
\setcounter{figure}{0} 
\subsection*{DFT calculation}

\begin{figure}[H]
\centering
    \includegraphics[width=0.9\linewidth]{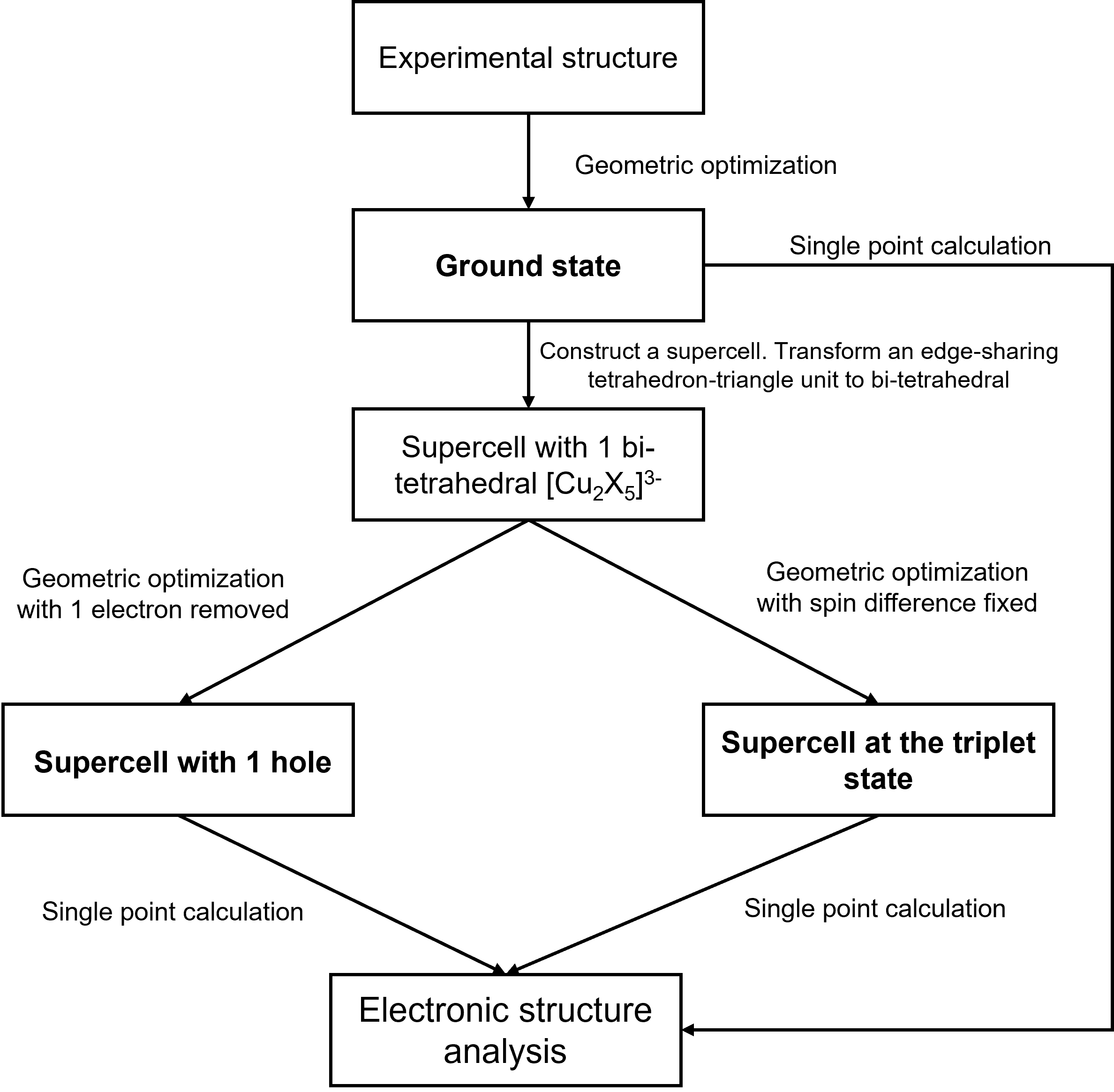}
    \caption{Flowchart for the DFT simulation of Cs$_3$Cu$_2$X$_5$. Ground state: DFT-optimized experimental structure, without specifying any electronic setting. Supercell with one hole: DFT optimized-supercell including one manually made bi-tetrahedral [Cu$_2$X$_5$]$^{3-}$ unit, with one electron removed in the total charge. Supercell at the triplet state: DFT optimized supercell including one manually made bi-tetrahedral [Cu$_2$X$_5$]$^{3-}$ unit, with the number difference between two spin channel being two.}
    \label{scheme}
\end{figure}
\begin{figure}[H]

\centering
    \includegraphics[width=0.9\linewidth]{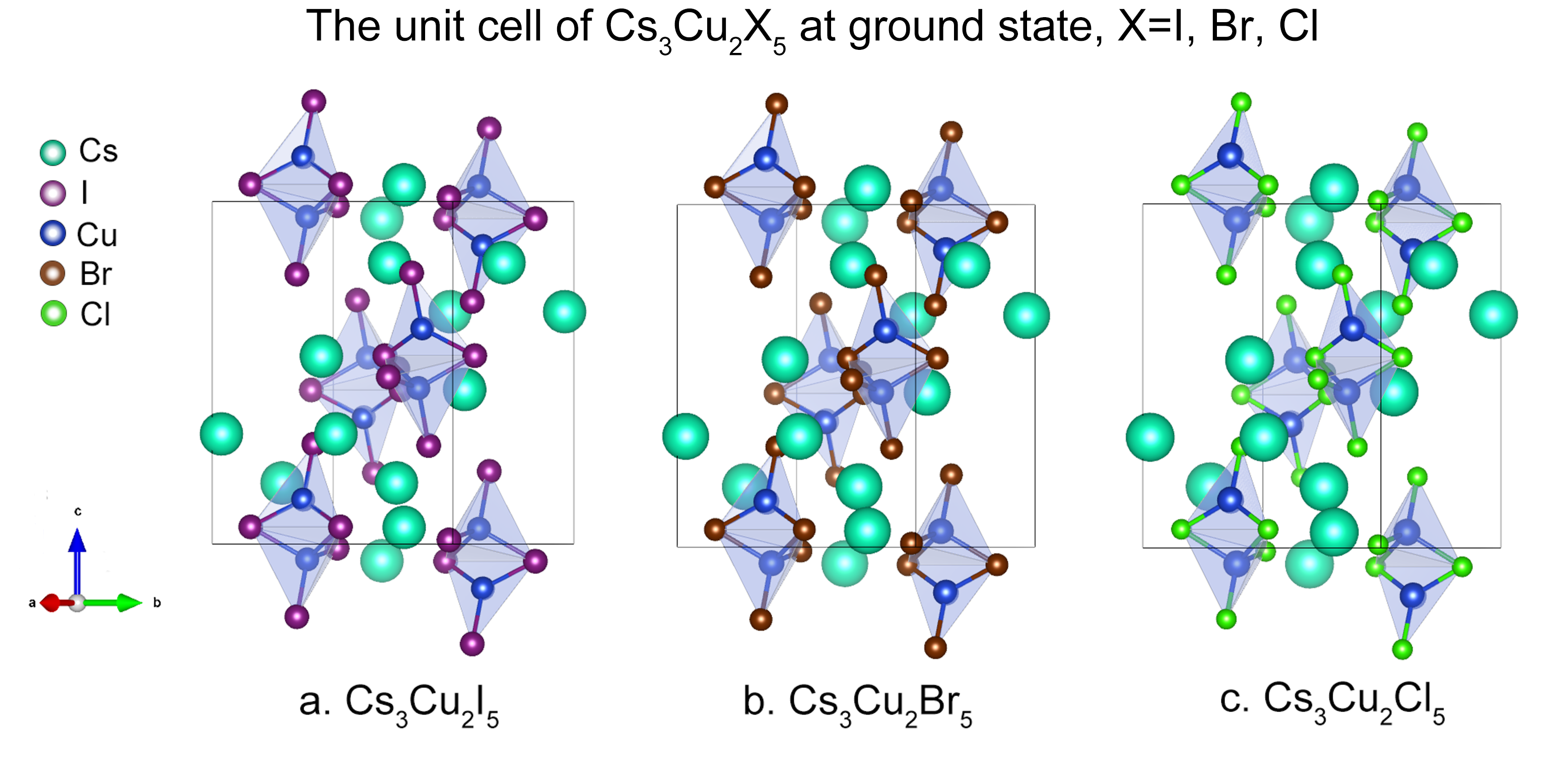}
    \caption{Simulated unit cells of 0D Cs$_3$Cu$_2$X$_5$ at ground states. One unit cell contains four [Cu$_2$X$_5$]$^3-$ blocks.}
    \label{si_ground_geo}
\end{figure}

\begin{figure}[H]
\centering
    \includegraphics[width=1\linewidth]{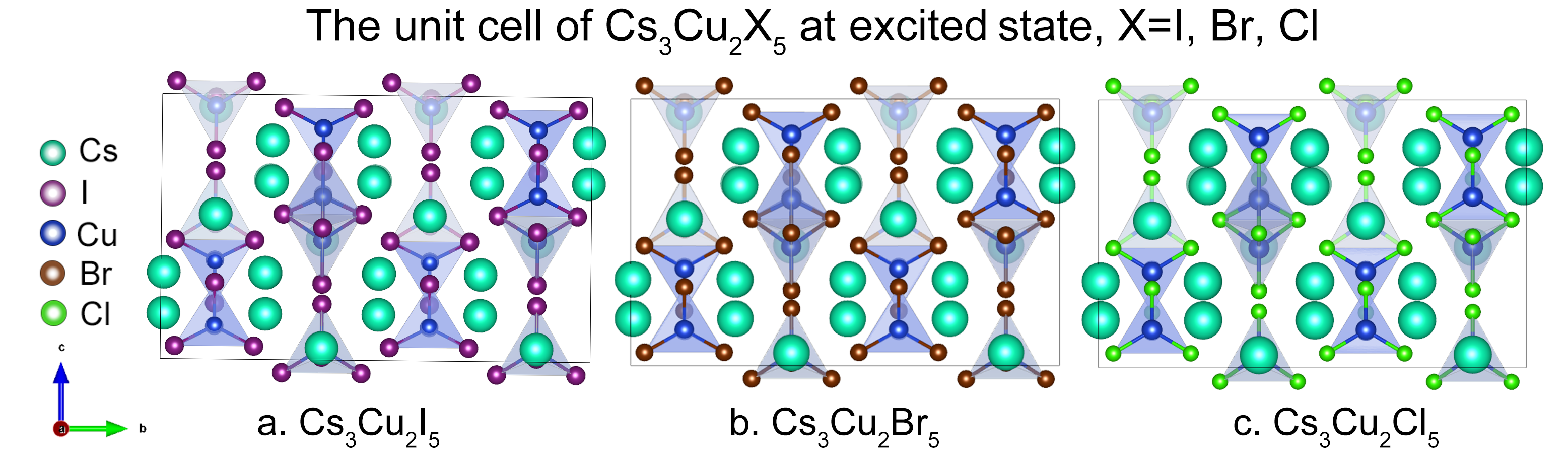}
    \caption{Simulated supercells of 0D Cs$_3$Cu$_2$X$_5$ at triplet states. One supercell contains 16 [Cu$_2$X$_5$]$^3-$ blocks.}
    \label{si_ex_geo1}
\end{figure}
\begin{figure}[H]
\centering
    \includegraphics[width=0.4\linewidth]{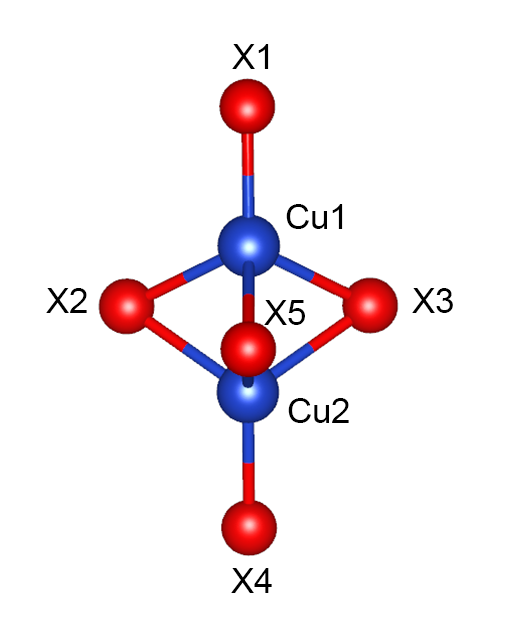}
    \caption{Schematic structure of 0D [Cu$_2$X$_5$]$^{3-}$ cluster with atom labels corresponding to the bond lengths in Table \ref{ground_bond} and \ref{triplet_bond}.}
    \label{label}
\end{figure}
\begin{table}[H]
    \centering
    \setlength{\tabcolsep}{10pt}        
    \setlength{\extrarowheight}{2pt}    
    \caption{Structural data of 0D $\text{Cs}_3\text{Cu}_2\text{X}_5$ at ground states. All unit cell are orthorhombic, with space group Pnma. The bond length with evident changes after excitation is marked in bold}
    \label{ground_bond}
    \begin{tabular}{ >{\centering\arraybackslash}m{4cm} | c|c|c } 
        \toprule 
        \multicolumn{4}{c}{\textbf{Unit: \AA}} \\
        \midrule

        \multicolumn{1}{>{\centering\arraybackslash}m{4cm}|}{\textbf{Lattice parameters}} & \textbf{a} & \textbf{b} & \textbf{c} \\
        \midrule
        $\text{X}=\text{I}$ & 10.08 & 11.55 & 14.23 \\
        $\text{X}=\text{Br}$ & 9.40 & 10.83 & 13.46 \\
        $\text{X}=\text{Cl}$ & 9.06 & 10.37 & 12.98 \\
        \midrule 
        \multicolumn{1}{>{\centering\arraybackslash}m{4cm}|}{\textbf{Distance between $\text{Cu}^{+}$ and $\text{X}^{-}$ in the $[\text{Cu}_2\text{X}_5]^{2-}$}} & \textbf{$\text{X}=\text{I}$} & \textbf{$\text{X}=\text{Br}$} & \textbf{$\text{X}=\text{Cl}$} \\
        \midrule
        $\text{Cu}1\text{-}\text{X}1$ & 2.50 & 2.31 & 2.17 \\
        $\text{Cu}1\text{-}\text{X}2$ & 2.54 & 2.31 & 2.23 \\
        $\text{Cu}1\text{-}\text{X}3$ & 2.54 & 2.36 & 2.23 \\
        \textbf{$\text{Cu}1\text{-}\text{X}5$} & \textbf{3.42} & \textbf{3.34} & \textbf{3.28} \\
        $\text{Cu}2\text{-}\text{X}2$ & 2.72 & 2.58 & 2.46 \\
        $\text{Cu}2\text{-}\text{X}3$ & 2.72 & 2.58 & 2.46 \\
        $\text{Cu}2\text{-}\text{X}4$ & 2.55 & 2.36 & 2.22 \\
        $\text{Cu}2\text{-}\text{X}5$ & 2.58 & 2.36 & 2.23 \\
        \midrule 
        \multicolumn{1}{>{\centering\arraybackslash}m{4cm}|}{\textbf{$\text{Cu}1\text{-}\text{Cu}2$ distance}} & \textbf{$\text{X}=\text{I}$} & \textbf{$\text{X}=\text{Br}$} & \textbf{$\text{X}=\text{Cl}$} \\
        \midrule
        $\text{Cu}1\text{-}\text{Cu}2$ & 2.48 & 2.45 & 2.45 \\
        \bottomrule 
    \end{tabular}
\end{table}

\begin{table}[H]
    \centering
    \setlength{\tabcolsep}{10pt}        
    \setlength{\extrarowheight}{2pt}    
    \caption{Structural data of 0D $\text{Cs}_3\text{Cu}_2\text{X}_5$ at triplet states. All unit cell are orthorhombic, with space group Pnma. The bond length with evident changes after excitation is marked in bold}
    \label{triplet_bond}
    \begin{tabular}{ >{\centering\arraybackslash}m{4cm} | c|c|c } 
        \toprule 
        \multicolumn{4}{c}{\textbf{Unit: \AA}} \\
        \midrule 
        \multicolumn{1}{>{\centering\arraybackslash}m{4cm}|}{\textbf{Lattice parameters}} & \textbf{a} & \textbf{b} & \textbf{c} \\
        \midrule
        $\text{X}=\text{I}$ & 10.08 & 11.55 & 14.23 \\
        $\text{X}=\text{Br}$ & 9.40 & 10.83 & 13.46 \\
        $\text{X}=\text{Cl}$ & 9.06 & 10.37 & 12.98 \\
        \midrule 
        \multicolumn{1}{>{\centering\arraybackslash}m{4cm}|}{\textbf{Distance between $\text{Cu}^{+}$ and $\text{X}^{-}$ in the $[\text{Cu}_2\text{X}_5]^{2-}$}} & \textbf{$\text{X}=\text{I}$} & \textbf{$\text{X}=\text{Br}$} & \textbf{$\text{X}=\text{Cl}$} \\
        \midrule
        $\text{Cu}1\text{-}\text{X}1$ & 2.50 & 2.31 & 2.16 \\
        $\text{Cu}1\text{-}\text{X}2$ & 2.63 & 2.43 & 2.28 \\
        $\text{Cu}1\text{-}\text{X}3$ & 2.63 & 2.42 & 2.28 \\
        \textbf{$\text{Cu}1\text{-}\text{X}5$} & \textbf{2.81} & \textbf{2.70} & \textbf{2.57} \\
        $\text{Cu}2\text{-}\text{X}2$ & 2.71 & 2.54 & 2.41 \\
        $\text{Cu}2\text{-}\text{X}3$ & 2.71 & 2.56 & 2.41 \\
        $\text{Cu}2\text{-}\text{X}4$ & 2.51 & 2.32 & 2.17 \\
        $\text{Cu}2\text{-}\text{X}5$ & 2.63 & 2.44 & 2.30 \\
        \midrule 
        \multicolumn{1}{>{\centering\arraybackslash}m{4cm}|}{\textbf{$\text{Cu}1\text{-}\text{Cu}2$ distance}} & \textbf{$\text{X}=\text{I}$} & \textbf{$\text{X}=\text{Br}$} & \textbf{$\text{X}=\text{Cl}$} \\
        \midrule
        $\text{Cu}1\text{-}\text{Cu}2$ & 2.37 & 2.37 & 2.35 \\
        \bottomrule 
    \end{tabular}
\end{table}
\begin{figure}[H]
\centering
    \includegraphics[width=0.7\linewidth]{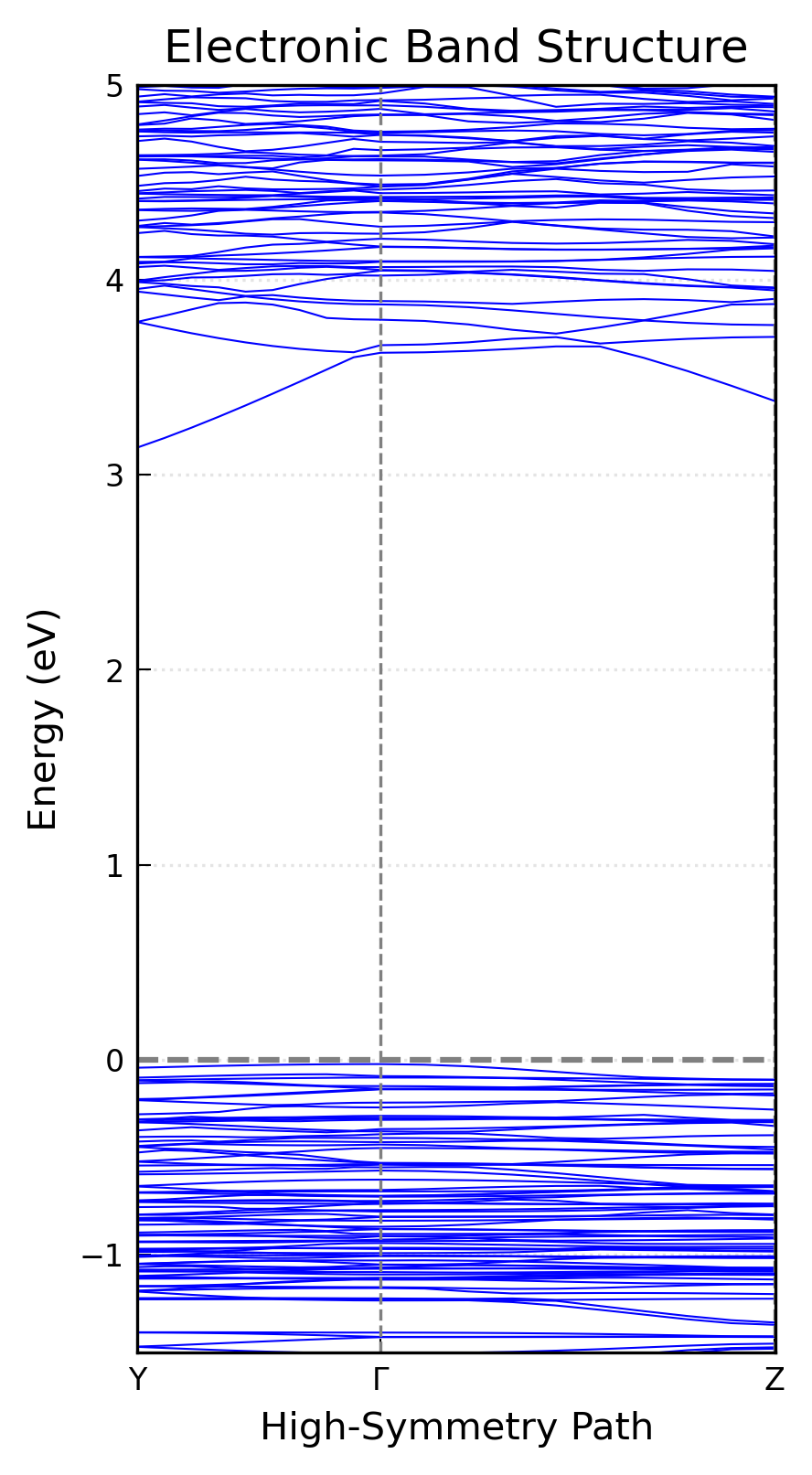}
    \caption{Band structure of Cs$_3$Cu$_2$I$_5$ at ground state. The Fermi level is relocated at zero.}
    \label{i_band}
\end{figure}
\begin{figure}[H]
\centering
    \includegraphics[width=1\linewidth]{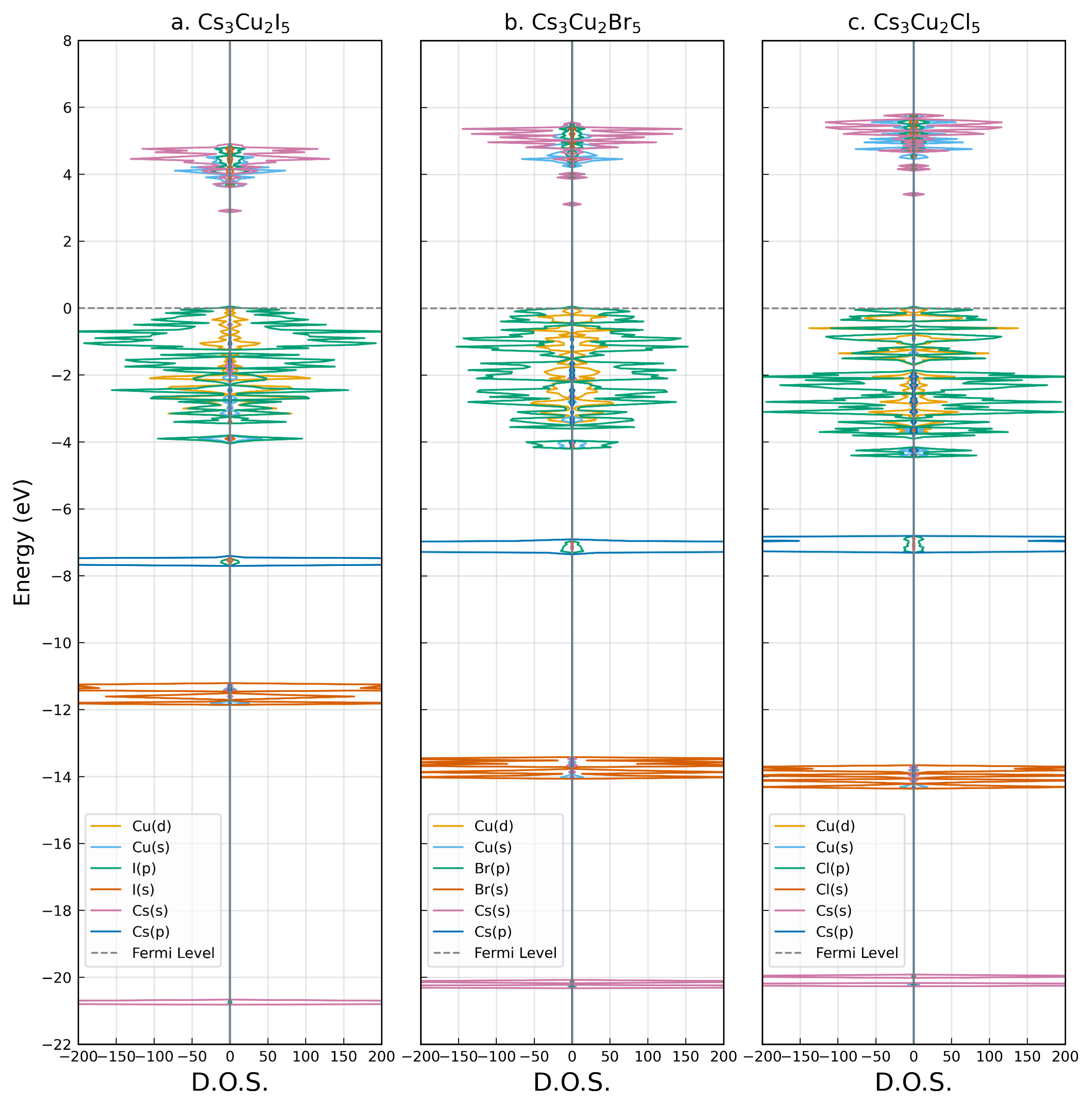}
    \caption{Density of states 0D Cs$_3$Cu$_2$
    X$_5$ at ground states. }
    \label{si_dos}
\end{figure}
\begin{figure}[H]
\centering
    \includegraphics[width=0.6\linewidth]{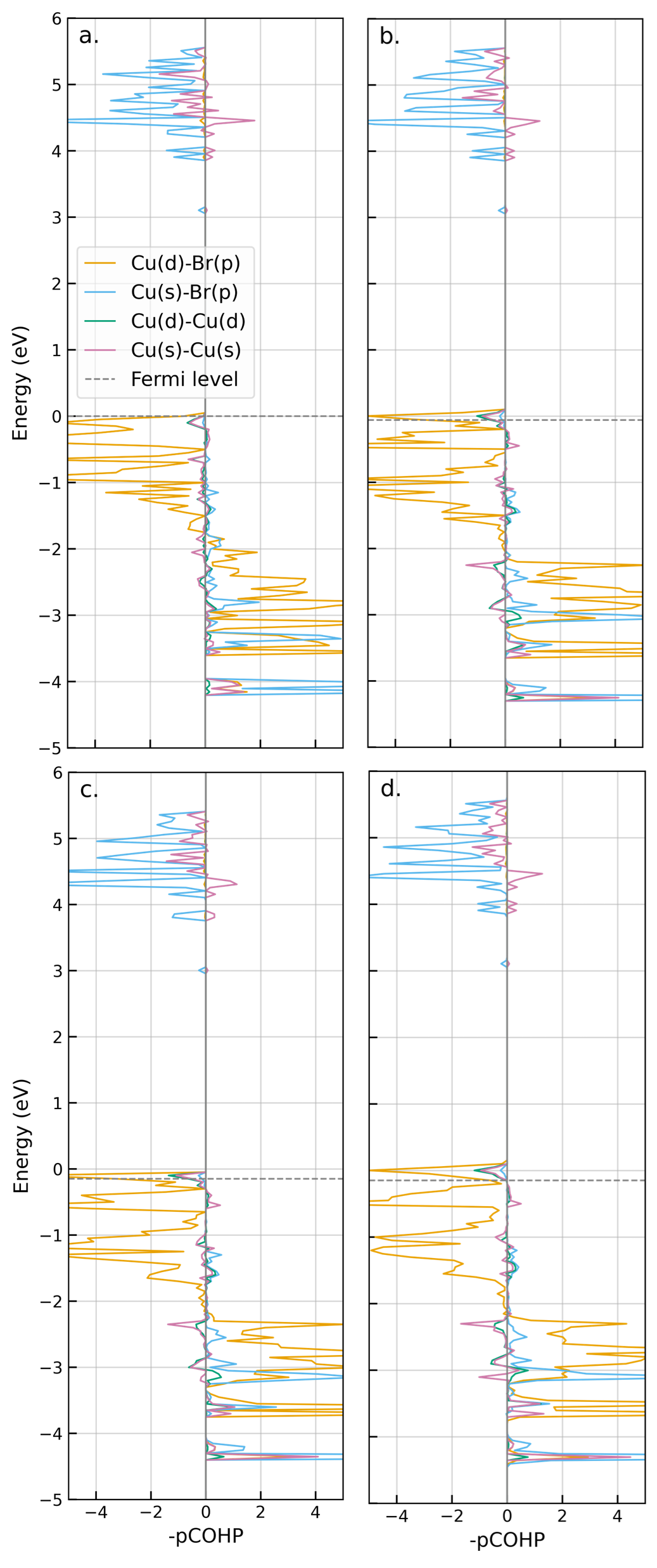}
    \caption{Partial COHP of 0D Cs$_3$Cu$_2$Br$_5$ at (a) ground state, (b) one hole added, and the hole channels of the (c) singlet and (d) triplet state}
    \label{si_br_cohp}
\end{figure}
\begin{figure}[H]
\centering
    \includegraphics[width=0.6\linewidth]{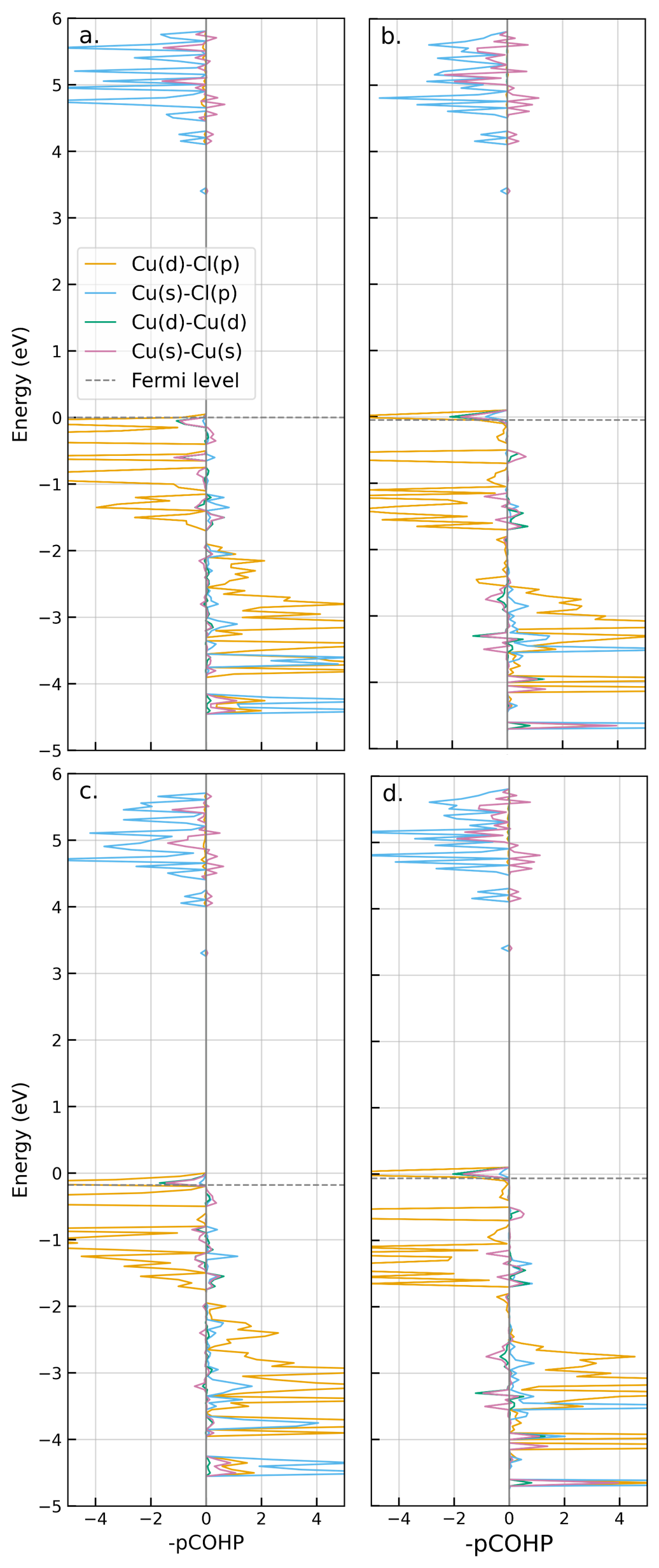}
    \caption{Partial COHP of 0D Cs$_3$Cu$_2$Cl$_5$ at (a) ground state, (b) one hole added, and the hole channels of the (c) singlet and (d) triplet state}
    \label{si_cl_cohp}
\end{figure}
\end{document}